\documentclass[aps,prl,reprint,superscriptaddress,floatfix,amsmath]{revtex4-2} 

\usepackage{ulem} 
\usepackage{bm} 
\usepackage{amssymb}
\usepackage{amsfonts} 
\usepackage{amsmath} 

\usepackage{graphicx} 

\usepackage{xcolor} 

\usepackage[pdftex]{hyperref}  
\hypersetup{
	colorlinks=true,
	breaklinks=true,
	allcolors=blue,
	pdfdisplaydoctitle=true,
	pdfauthor = {Alexander Johannes Edward Kreil}
} 

\usepackage[utf8]{inputenc} 

\definecolor{g-blue}{rgb}{0.83,0.95,1}
\definecolor{g-yellow}{rgb}{1,1,0.7}
\definecolor{g-green}{rgb}{0.9,1,0.9}
\definecolor{green}{rgb}{0,0.6,0}
\definecolor{cyan}{rgb}{0,0.7,0.7}
\definecolor{black}{rgb}{0,0,0}
\definecolor{grey}{rgb}{0.4,0.4,0.4}
\definecolor{nature-blue}{rgb}{0.0,0.200,0.500}

\def \ed {\end{document}}
\def\Fbox#1{\vskip1ex\hbox to 8.5cm{\hfil\fboxsep0.3cm\fbox{%
		\parbox{8.0cm}{#1}}\hfil}\vskip1ex\noindent}  

\def\be{\begin{equation}}
\def\ee{\end{equation}}
\def\bea{\begin{eqnarray}}
\def\eea{\end{eqnarray}}
\def\bse{\begin{subequations}}
	\def\ese{\end{subequations}}

\let \= \equiv  
\let\*\cdot 
\let\^\widehat 
\let\-\overline
  
\def\ort#1{\^{\bf{#1}}}

\def\x{\ort x} \def\y{\ort y} 
 \def\1{\bm1} 

\def\<{\left\langle}    \def\>{\right\rangle}
\def\({\left(}          \def\){\right)}
\def \[ {\left [} \def \] {\right ]}

\newcommand{\Eq}[1]{Eq.\,(\ref{#1})}
\newcommand{\Eqs}[1]{Eqs.\,(\ref{#1})}
\newcommand{\Fig}[1]{Fig.\,\ref{#1}}
\newcommand{\Figs}[1]{Figs.\,\ref{#1}}

\renewcommand{\sb}[1]{_{\text {#1}}}  
\renewcommand{\sp}[1]{^{\text {#1}}}  
\def\Sb#1{_{\scriptscriptstyle\rm{#1}}}

\begin{document}
	
	\title{Experimental observation of Josephson oscillations in a room-temperature Bose--Einstein magnon condensate}
	
	\author{Alexander~J.~E.~Kreil}	
	\affiliation{Fachbereich Physik and Landesforschungszentrum OPTIMAS, Technische Universit\"at Kaiserslautern, 67663 Kaiserslautern, Germany \looseness=-1}
	
	\author{Halyna~Yu.~Musiienko-Shmarova}
	\affiliation{Fachbereich Physik and Landesforschungszentrum OPTIMAS, Technische Universit\"at Kaiserslautern, 67663 Kaiserslautern, Germany \looseness=-1}
	
	\author{Pascal~Frey}	
	\affiliation{Fachbereich Physik and Landesforschungszentrum OPTIMAS, Technische Universit\"at Kaiserslautern, 67663 Kaiserslautern, Germany \looseness=-1}
	
	\author{Anna~Pomyalov}
	\affiliation{Department of Chemical and Biological Physics, Weizmann Institute of Science, Rehovot 76100, Israel \looseness=-1}
	
	\author{Victor~S.~L'vov}
	\affiliation{Department of Chemical and Biological Physics, Weizmann Institute of Science, Rehovot 76100, Israel \looseness=-1}
	
	\author{Gennadii~A.~Melkov}
	\affiliation{Faculty of Radiophysics, Electronics, and Computer Systems, Taras Shevchenko National University of Kyiv, Kyiv 01601, Ukraine \looseness=-1}

	\author{Alexander~A.~Serga}
	\email{serga@physik.uni-kl.de}
	\affiliation{Fachbereich Physik and Landesforschungszentrum OPTIMAS, Technische Universit\"at Kaiserslautern, 67663 Kaiserslautern, Germany \looseness=-1}
	
	\author{Burkard~Hillebrands}
	\affiliation{Fachbereich Physik and Landesforschungszentrum OPTIMAS, Technische Universit\"at Kaiserslautern, 67663 Kaiserslautern, Germany \looseness=-1}

\begin{abstract}

The alternating current (AC) Josephson effect in a time-independent spatially-inhomogeneous setting is manifested by the occurrence of Josephson oscillations---periodic macroscopic phase-induced collective motions of the quantum condensate. So far, this phenomenon was observed at cryogenic temperatures in superconductors, in superfluid helium, and in Bose--Einstein condensates (BECs) of trapped atoms. Here, we report on the discovery of the AC Josephson effect in a magnon BEC carried by a room-temperature ferrimagnetic film. The BEC is formed in a parametrically populated magnon gas in the spatial vicinity of a magnetic trench created by a DC electric current. The appearance of the Josephson effect is manifested by oscillations of the magnon BEC density in the trench, caused by a coherent phase shift between this BEC and the BEC in the nearby regions. Our findings advance the physics of room-temperature macroscopic quantum phenomena and will allow for their application for data processing in magnon spintronics devices.	 

\end{abstract}
	
\maketitle 

\section{I. Introduction}
\vspace{-3mm}
	
The  ac~Josephson effect \cite{Josephson1962} is well known as a rapidly oscillating current that appears between two weakly coupled superconductors subject to an external DC voltage. Soon after the first experimental proof of principle was made \cite{Anderson1963}, the Josephson effect was used in various applications such as voltage standards, ultrasensitive magnetic field sensors, and quantum computing \cite{Orlando1999,Cerletti2005,Arute2019}. It was intensively studied in various configurations \cite{Likharev1979, Barone1982}. 
A similar behavior has been observed in bosonic systems such as Bose--Einstein condensates (BECs) in superfluid $^3$He \cite{Borovik-Romanov1988, Pereverzev1997, Backhaus1997, Davis2002, Autti2020}, $^4$He \cite{Sukhatme2001}, in weakly interacting atomic Bose gases \cite{Anderson1995, Davis1995, Bradley1995, Albiez2005, Levy2007}, and in exciton-polariton condensates \cite{Shelykh2008, Sarchi2008, Lagoudakis2010, Abbarchi2013}. Recently, magnon supercurrents were observed in room-temperature ferrimagnetic films \cite{Bozhko2016}, and the existence of the related magnon Josephson effect was theoretically predicted \cite{Schilling2012, Troncoso2014, Nakata2014, Nakata2015, Nakata2017}. 
	
In this paper, we present the experimental discovery and a theoretical analysis of Josephson oscillations in a room-temperature Yttrium Iron Garnet \cite{Cherepanov1993} (Y$_3$Fe$_5$O$_{12}$, YIG) ferrimagnetic film, by exploring the spatio-temporal dynamics of a magnon BEC \cite{Demokritov2006} prepared by microwave parametric pumping \cite{Rezende1990, Lvov1994, Gurevich1996}. 
 
In general, AC Josephson oscillations appear in a system of two connected BECs with different energies, and this difference sets the oscillation frequency.
In the well-known scenario of two BECs formed by Cooper pairs in superconductors, which are weakly linked via a tunneling  barrier between them, this energy difference is determined by the electric field's potential difference across the barrier. The tunneling  barrier preserves the potential difference and, at the same time, ensures a current flow between the superconductors. 
 
In the case of a magnon BEC, the energy of magnons depends on the bias magnetic field. Thus, to observe the magnon AC Josephson oscillations, it is sufficient to create a magnetic field inhomogeneity to obtain the energy difference between two BECs, leaving them under well-chosen conditions with a reduced coupling between them. Such a magnetic field inhomogeneity can be prepared without any tunneling barrier.
This can be realized, for instance, conveniently by an electric wire positioned across a magnetic sample and carrying a DC electric current. It produces locally a magnetic field and thus creates areas containing BECs with different energies.
Our work explores such a setting.
Of course, this in no way negates the possibility of implementing the magnon Josephson effect in its canonical form, with a tunneling barrier, as proposed, for instance, in Refs.\,\cite{Schilling2012,Troncoso2014,Nakata2014,Nakata2015,Nakata2017}. \looseness=-1
 
\vspace{-3mm}
\section{II. Experiment}
\vspace{-3mm}
\subsection{1. Idea of experiment}
\vspace{-3mm}
 
\begin{figure}[t]\textbf{}
	\includegraphics[width=1\columnwidth]{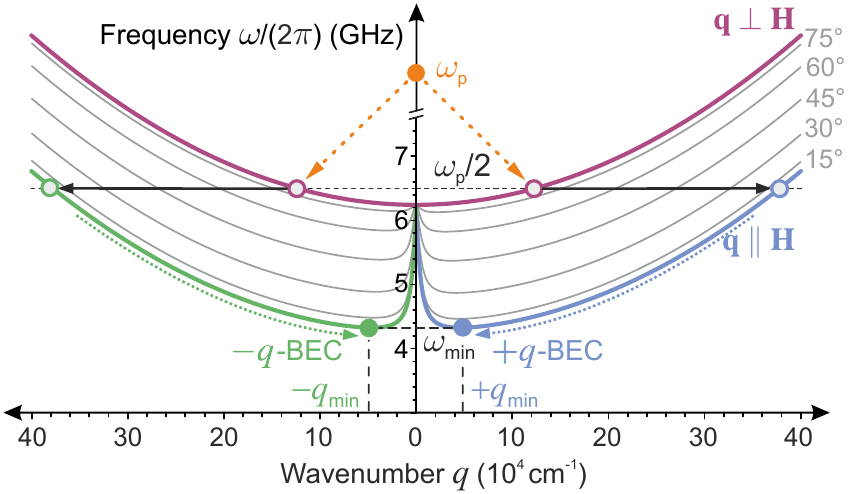}
	\caption{
		\label{f:1} 
		Spectrum of the fundamental magnon mode in a 5.1-$\mu$m-thick YIG film magnetized in plane by a bias magnetic field $H=1520$\,Oe, shown for the wavevector $\bm{q} \parallel \bm{H}$ (lower part of the spectrum, blue and green curves), for $ \bm{q} \perp \bm{H} $ (upper part, magenta curves), and also for several intermediate directions of wavevectors (grey curves). The orange arrows illustrate the magnon injection process by means of parallel parametric pumping. The horizontal black arrows illustrate scattering of the pumped magnons into secondary magnon states. The blue and green dotted arrows symbolize the flow of thermalizing magnons to the bottom of the spectrum. The blue and green dots indicate the positions of the frequency minima $\omega\sb{min}(+\bm{q}\sb{min})$ and $\omega\sb{min}(-\bm{q}\sb{min}$) occupied by the $+q$- and $-q$-BECs of magnons.
	}    
\end{figure}

In our experiments we take into account two facts: 
 
(i) The magnon frequency spectrum in an in-plane magnetized magnetic film has two equivalent minima $\omega\sb{min}=\omega(\pm \bm{q}_\mathrm{min})$ at symmetric values of the wavevector.
Therefore, the magnon BEC consists of two components, named subsequently as $+q$-BEC and $-q$-BEC, localized in the $\bm{q}$-space around $+\bm{q}_\mathrm{min}$ and $-\bm{q}_\mathrm{min}$, respectively, see \Fig{f:1}. \looseness=-1
 
\begin{figure*}  		 
	\includegraphics[width=1\textwidth]{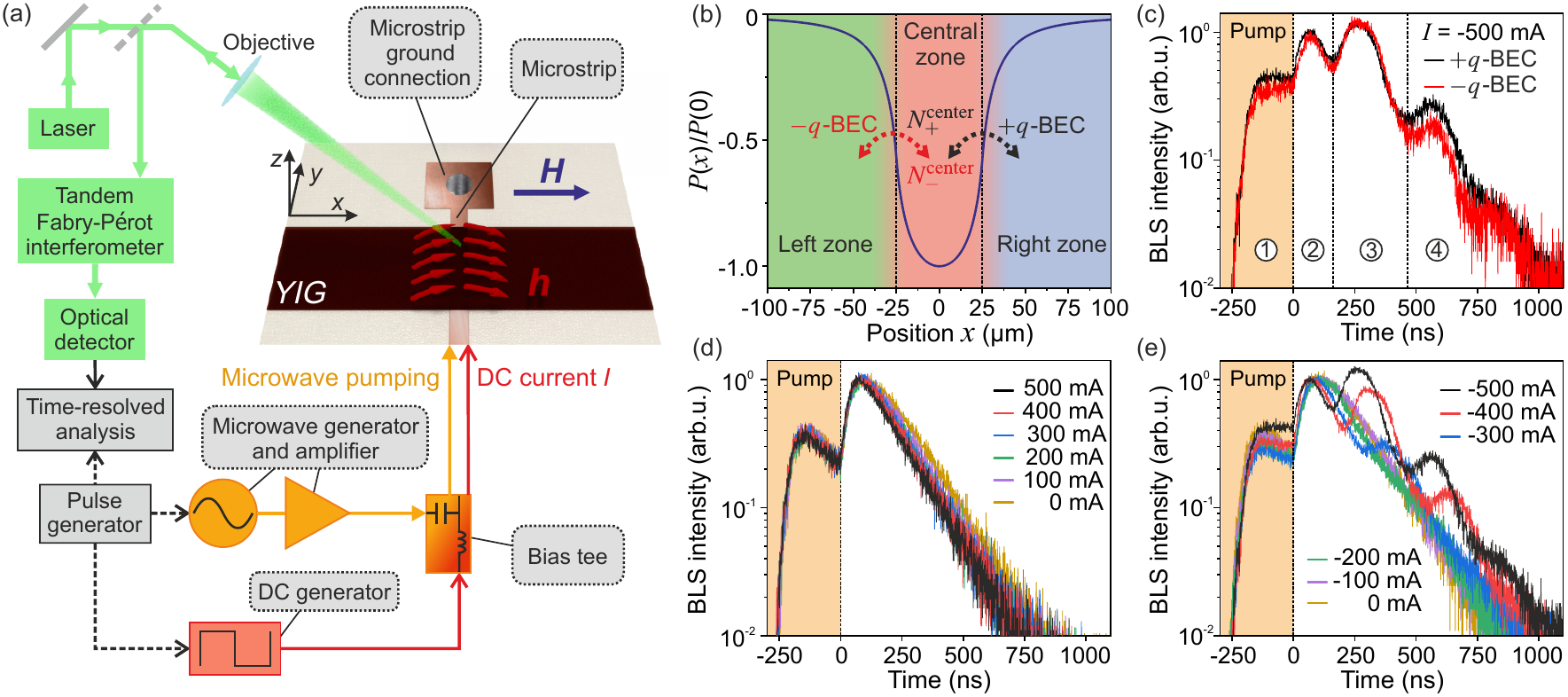}
	\caption{
		\label{f:2}
		Experimental setup and Josephson oscillations. (a) Sketch of the experimental setup. A YIG-film magnetized in-plane by a  magnetic field $\bm{H}$ is placed on top of a 50\,$\mu$m-wide microstrip. The red and orange branches represent the circuits for the generation of DC and microwave currents. The red arrows illustrate the magnetic field $\bm{h}$ induced by a DC current. The green part describes a tool for optical detection of magnons by means of Brillouin light scattering spectroscopy (BLS). A pulse generator triggers DC and microwave generation, and synchronizes the time-resolved BLS analysis. 
		(b) The calculated profile of a potential trench $P(x)$ controlling the energy of BECs in different zones of the sample. The bidirectional dashed arrows indicate two alternating magnon supercurrents related to the oscillations of $+q$-BEC (black arrow) and $-q$-BEC (red arrow) densities $N_+\sp{center}$ and $N_-\sp{center}$, observed in the central zone of the potential trench.
		Panels (c), (d), (e) present the BLS intensity in the central zone, which is proportional to the magnon density at the bottom of spin-wave spectrum, versus time. (c) Josephson oscillations in two systems, $+q$-BEC (black line) and $-q$-BEC (red line). Time intervals 1-4 in (c) mark: 1 -- the time, when magnons are excited by parametric pumping, 2 -- the time interval during which the BEC forms, 3 -- the time interval during which the first oscillation appears, and 4 -- the rest of the observation time. 
		Panels (d) and (e) present the summarized dynamics of two BEC systems $(N_+\sp{center}+N_-\sp{center})$ for different bias DC currents:
		(d) Potential ridge formed by DC currents from 0 to $+500\,$mA. 
		(e) Potential trench formed by DC currents from 0 to $-500\,$mA.  
	}
\end{figure*}

(ii) Using a space-inhomogeneous pumping mechanism by a microstrip antenna shown in \Fig{f:2}(a), we can populate $\pm q$-BECs such that their densities in the physical space, $N_+(x)$ and $N_-(x)$, become different.
Using Brillouin Light Scattering (BLS) spectroscopy (see \Fig{f:2}(a) and Appendix) we show that the central zone of the sample [under the microstrip, see \Fig{f:2}(b)] is populated almost equally  $N_+(x)\simeq N_-(x)\simeq N\sp{center}$, while in the left zone $N_-\sp{left} > N_+\sp{left}$ and in the right zone $N_+\sp{right} > N_-\sp{right}$.
 
The central idea of our experiment is to create a spatially-inhomogeneous potential energy profile for condensed magnons by subjecting the magnetic film to a time-independent space-inhomogeneous magnetic field $\bm{h}(x)$ added to the uniform bias magnetic field $\bm{H}$. 

Since in our experiment the magnetic field $\bm{h}(x)$ is induced by a DC electric current flowing through the pump microstrip [see \Fig{f:2}(a)], this energy profile has---depending on the polarity of the current---the shape of a trench or ridge across the microstrip. This profile is extended along the microstrip. The potential energy profile 
	\begin{equation}\label{gamma}
	P(x)\approx \gamma h_x(x) \= \Omega_0\, f(x)\,,  
	\end{equation}
calculated for our experimental conditions in the direction perpendicular to the microstrip, is shown in \Fig{f:2}(b) for the trench case. In \Eq{gamma}, $\gamma$ is the gyromagnetic ratio, $f(x)$ is the normalized energy profile, defined such that $f(0)=1$, and frequency $\Omega_0$ is the profile magnitude: $\Omega_0>0$ for the ridge and  $\Omega_0 <0$ for the trench. In our experiment, $|\Omega_0| =2\pi \cdot 1.6\,\cdot 10^5 I\,$rad/s, if the value $I$ is the DC current measured in milliampere (see Appendix).
 
In such a configuration, for each of the condensates, $+q$-BEC and $-q$-BEC, we create two Josephson junctions, which are located between the central zone and the two side zones.
The difference in the potential energies leads to the phase shift between different spatial parts of the same magnon condensate ($-q$-BEC or $+q$-BEC): at the bottom of the trench and in the side zones of the sample. 
This phase shift propels AC magnon supercurrents between the central zone and the left and the right zones.
We detect these AC supercurrents as the Josephson oscillations of the $+q$-BECs and $-q$-BECs occupations in the middle of the trench [see \Figs{f:2}(c) and \ref{f:2}(e)]. As seen in \ref{f:2}(d), there are no oscillations when a potential ridge is created in the central zone instead of the trench. \looseness=-1

\vspace{-3mm}
\subsection{2. Experimental setup}
\vspace{-3mm}
		
Magnons are parametrically pumped \cite{Cherepanov1993, Lvov1994, Gurevich1996} in an in-plane magnetized $5.1\,\mu$m-thick YIG-film grown on a Gadolinium Gallium Garnet substrate. During pumping, microwave photons with frequency $\omega\sb p$ decay into two magnons with frequency $\omega\sb p/2$ and wavevectors $\pm \bm{q}$, mainly with $\bm{q} \perp \bm{H}$ \cite{Serga2012, Neumann2009}, see \Fig{f:1}.

As a result of multiple intermagnon elastic scattering events with two input and two output magnons, the parametrically pumped magnons thermalize, populating the lower energy states. This process leads to an increase in the chemical potential of the magnon gas. When the value of the chemical potential reaches the bottom of the magnon spectrum, coherent magnon condensates, i.e. the $-q$-BEC and the $+q$-BEC are formed there \cite{Demokritov2006}. Note also that the thermalization process involves isotropization of the wavevector directions and, consequently, leads to the appearance of magnons that have wavevector components parallel to the magnetic field $\bm{H}$. An increase in the number of such magnons allows the formation of BECs with $\pm \bm{q}\sb{min} \parallel \bm{H}$.

A schematic view of the experiment is shown in \Fig{f:2}(a). Microwave pulses with the frequency $\omega\sb p/(2\pi)= 13$\,GHz and a duration of $\tau\sb{p} = 300\,$ns feed a $50\,\mu$m-wide microstrip, inducing a microwave pumping field. The microstrip is fabricated on a dielectric substrate, and its end is connected to the ground plate on the backside of the substrate, creating an electrical shortcut in the electric circuit. The YIG-film sample is positioned on top of the microstrip.

Unlike in our previous work, where the magnon supercurrent was induced by a thermal inhomogeneity in a hot laser spot \cite{Bozhko2016,Bozhko2019}, in this work we used the DC electric current for producing a time-independent inhomogeneous magnetic field $\bm{h}(x)$.

To detect the magnons from the bottom of the spectrum, conventional wavevector- and time-resolved BLS was used \cite{Buettner2000, Sandweg2010, Serga2012}. The BEC-related BLS signal was detected around $\omega_\mathrm{min}$ (see Fig.\,\ref{f:1}). 
The value of an incident angle of the laser light was adjusted to detect the magnons of wavenumbers $\pm q_\mathrm{min}$. For more details on the experimental methods, see Appendix.

\vspace{-3mm}
\subsection{3. Experimental results}
\vspace{-3mm}
	
\begin{figure}[t]
	\includegraphics[width=1\columnwidth]{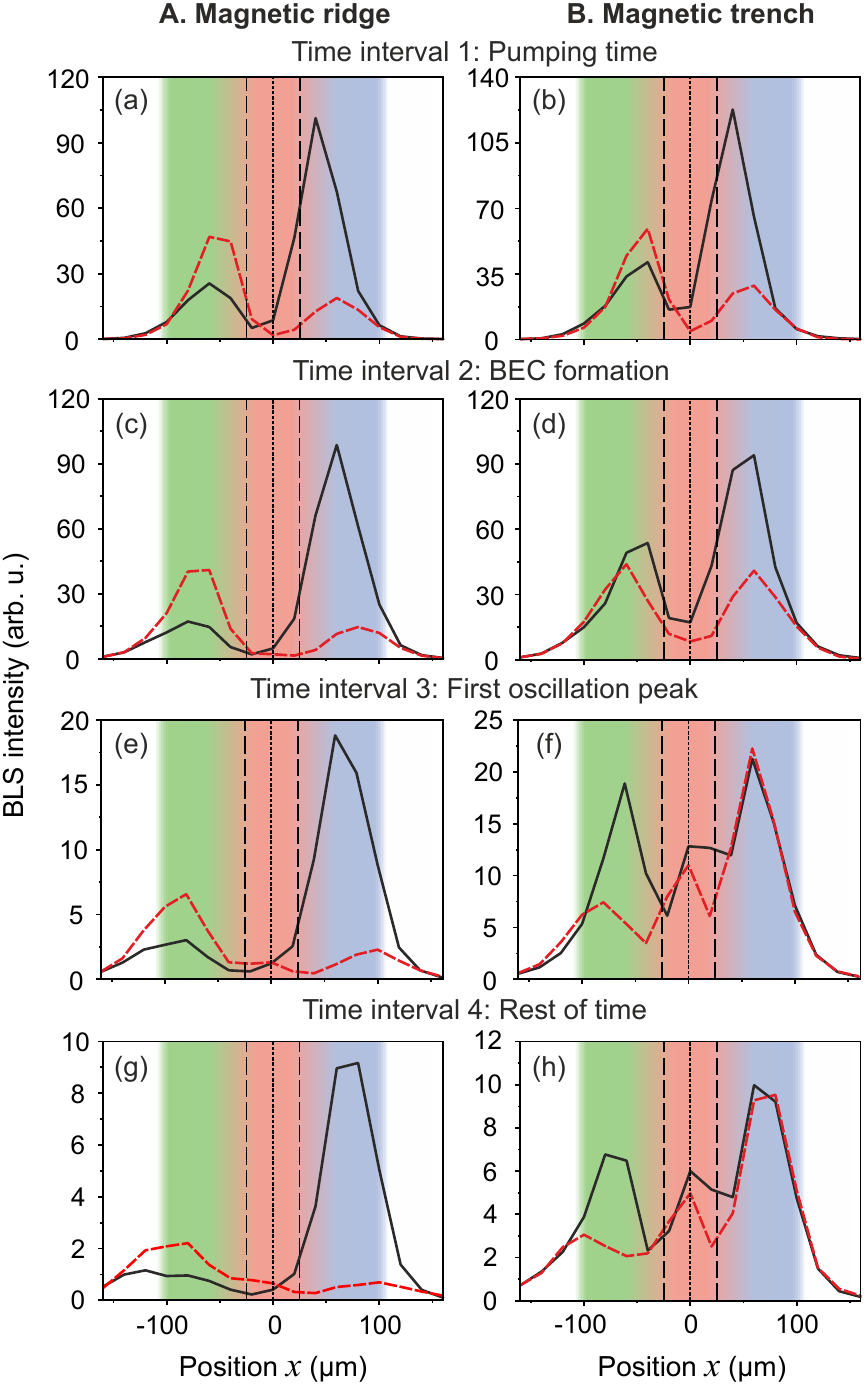}
	\caption{
		\label{f:3} 
		Time-dependence of spatial magnon distributions in different potential profiles. Distribution of magnons perpendicular to the antenna. Rows 1, 2, 3 and 4 show integrated data over different time intervals of the time-resolved measurement data, as shown in \Fig{f:2}(c). The intensity of the $+q$-BEC- and $-q$-BEC are shown by black solid and red dashed lines, respectively. The difference in the intensities of these condensates refers to the difference in the efficiency of the Stokes and anti-Stokes scattering processes (see Appendix). The first column (A) shows data for the magnetic ridge with the applied current $I= 500\,$mA, while the second column (B) corresponds to the magnetic trench with the applied current $I=-500\,$mA. Note the change in the BLS intensity scale.
	}
\end{figure}

Figure\,\ref{f:3} shows profiles $N_+(x)$ of the $+q$-BEC (solid black lines) and $N_-(x)$ of the $-q$-BEC (dashed red lines) obtained by scanning the BLS signal in $\bm{{\^x}}$-direction, perpendicular to the microstrip. We see that these BECs are separated in space: the $+q$-BEC is mostly concentrated in the right (blue colored) zone of the sample, while the $-q$-BEC -- in the left (green colored) sample zone. As expected from the physical view-point, the amplitudes and the spatial distribution of these components in our experiments are indeed almost mirror-symmetric with respect to the $x=0$ line. 

To understand why $+q$-BECs and $-q$-BECs are mostly localized on different sides of the antenna, recall that at the initial stage of their thermalization, parametrically pumped magnons scatter to spectral positions with wavevectors $\bm{q}\sb{s}$ oriented at arbitrary angles relative to the bias magnetic field $\bm{H}$. The two horizontal arrows show an example of such scattering events in \Fig{f:1}. One observes that the secondary magnons with $\omega(\bm{q}\sb{s})=\omega\sb{s} /2 \gg \omega\sb{min}$ have a rather large group velocity 
$v_x(\bm{q}\sb{s})=\partial \omega(\bm{q})/\partial q_x$, which is positive  for $q_{\mathrm{s},x} > 0$  and negative  for $q_{\mathrm{s},x} < 0$. Therefore, during the further thermalization and condensation processes, the magnons with $q_{\mathrm{s},x} > 0$ are traveling to the right of the antenna, while the magnons with $q_{\mathrm{s},x} < 0$ move to the left. As a result, the $+q$-BECs and $-q$-BECs are formed separated and localized at different antenna sides, as is shown in \Fig{f:3}. \looseness=-1
	
In the geometry of our experiment, this spatial separation of the two magnon BECs is a robust and very general phenomenon that takes place in a wide range of pumping powers and in the range of $H$ from about 1500\,Oe to about 2100\,Oe. 
	
To see how the spatial separation of the $\pm q$-BECs $N_\pm(x,t)$ develops during the time of observation, in \Fig{f:3} we present the profiles $N_+(x,t)$ and $N_-(x,t)$  (by solid and dashed lines, respectively) for four characteristic time intervals of the magnon evolution, in four rows: the interval (1) covers the duration of the pumping [Figs.\,\ref{f:3}(a,\,b)], during interval (2) the BEC peaks are formed [Figs.\,\ref{f:3}(c,\,d)] the interval (3) corresponds to the time around the first oscillation peak [Figs.\,\ref{f:3}(e,\,f)] and the period (4) covers the rest of the magnon lifetime [Figs.\,\ref{f:3}(g,\,h)]. The evolution of the BEC's spatial distributions for different magnetic fields profiles is shown for two spatial arrangements: the potential ridge with a height $\Omega_0/(2\pi)$ of about $80$\,MHz (column A) and the potential trench of the same depth (column B). These frequency values correspond to the DC currents $I=\pm 500$\,mA. 
	
First of all, we see that during active pumping [row\,(1), Figs.\,\ref{f:3}(a,\,b)] the spatial distributions of both the $+q$-BEC and $-q$-BEC are practically the same for the magnetic ridge and the trench scenario. This results from the action of a very intensive pumping microwave magnetic field, which shakes the entire magnon frequency spectrum with an amplitude of about $500\,$MHz. This frequency shift corresponds to a microwave current amplitude of 3.22\,A at a pumping power of 400\,W. In addition, the interaction of a vast number of parametric magnons with the bottom magnons causes inelastic scattering of the latter over the spectrum. These effects prevent the formation of a coherent BEC state and, thus, of any supercurrent-related magnon dynamics during this time interval \cite{Serga2014,Bozhko2019,Kreil2019}.
	
The $+q$-BEC and $-q$-BEC spatial distributions shown in Column A in Fig.\,\ref{f:3} do not change after the pumping is turned off.
This is because the high magnetic ridge practically blocks the exchange of the BEC densities between the left and right zones, pushing condensed magnons away from the central zone \cite{Borisenko2020}. 

\begin{figure}[t]
	\includegraphics[width=1\columnwidth]{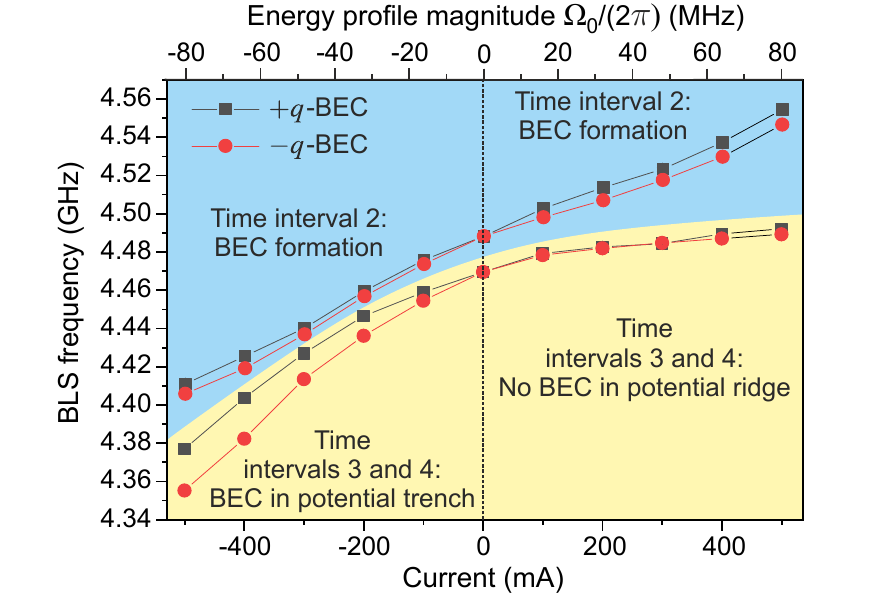}
	\caption{
		\label{f:4}
		Magnon frequency shift in the middle of the central zone of a potential profile.
		The center frequencies (found by a Gaussian fit) of the measured magnon spectra near $+q_\mathrm{min}$ (black lines and squares) and $-q_\mathrm{min}$ (red lines and dots) are shown as a function of the applied bias current $I$.
	}
\end{figure}
	
A completely different time evolution of the peaks is seen for the magnetic trench (Fig.\,\ref{f:3}, column B). Immediately after switching off the pumping, and still during the condensation process, the central zone becomes more populated compared to the magnetic ridge scenario discussed above. As the time progresses, the effect becomes even stronger. Such a behavior is quite expected: due to the lower magnon energy in the trench, the gaseous magnons fall into the trench populating the BEC states in the central zone.

The presence of the BECs in the trench and their absence in the ridge area is confirmed by the results of our BLS measurements of the magnon frequency in the central zone shown in Fig.\,\ref{f:4}.
During the BEC formation (blue part of the figure), the magnon frequency in the central zone depends almost linearly on the applied DC current. Later (yellow part of the figure), this dependence remains only for negative currents, which correspond to the case of the potential trench. No significant frequency shift is observed in the case of the potential ridge created by a positive DC current.

This behavior is quite expected. Indeed, our BLS setup has a spatial sensitivity of the order of the central zone width. In the case of the potential ridge, the BEC escapes the central zone. Therefore, the BLS measurements in the ridge center detect BEC magnons only at the ridge edges. This explains why the increase in the measured frequency in the yellow zone (time intervals 3 and 4) with a positive current is so small.

On the contrary, in the case of the trench, the fully-formed BEC falls further toward the trench bottom and causes a further decrease of the BEC frequency in the yellow zone at negative currents. Some differences in the $+q$- and $-q$-BEC frequencies may be due to the incomplete symmetry of the experimental settings.

The possibility for the BEC to be present in the central zone  is a necessary condition for the development of the Josephson oscillations, expressed in the periodic flow of magnons into the central zone and back into the right and left zones of the sample as symbolically shown in \Fig{f:2}(b) by the red and black dotted double arrows. The resulting oscillations of the BEC density, measured in the middle of the central zone ($x=0$), are clearly visible in \Fig{f:2}(c).

Note that the $\pm q$-BECs oscillate similarly. This allows us to ignore in the following analysis the difference in their behavior and to consider only the sum of these components $N\sp{center}(t)= N\sp{center}_+(t)+ N\sp{center}_-(t)$.

Figures\,\ref{f:2}(d) and \ref{f:2}(e) show the time evolution of $N\sp{center}(t)$ for the spatially-homogeneous geometry ($I=0$\,mA) and for the ridge and the trench geometries, respectively. As expected, the Josephson oscillations are observed only in the trench geometry.

\begin{figure}  
	\includegraphics[width=1\columnwidth]{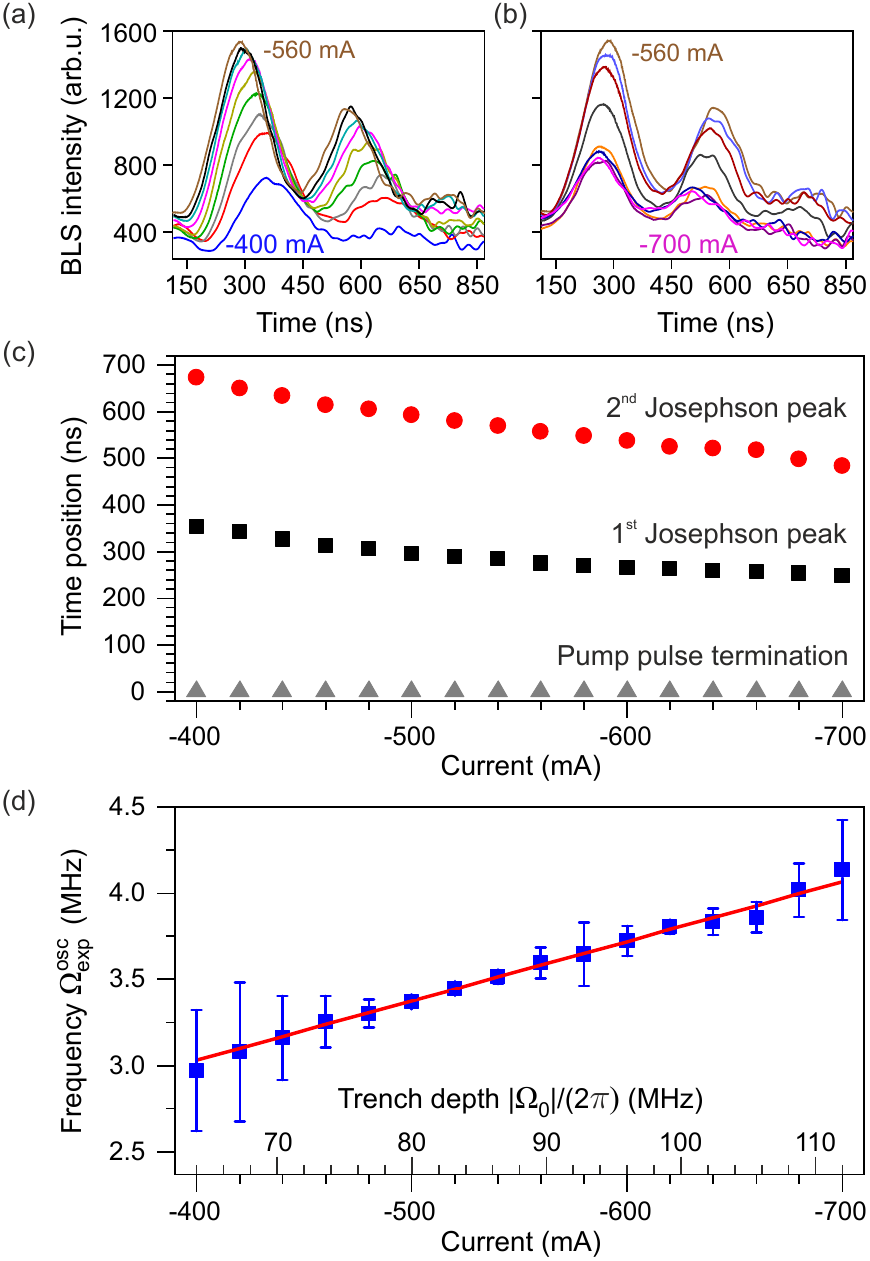}
	\caption{
		\label{f:5} 
		Determination of Josephson oscillation frequencies. 
		(a),(b) Time dependencies of the BLS intensity measured for various negative DC currents are shown after noise filtering and attenuation compensation. (a) The BLS intensity increases with the increase in current from $-400$\,mA to $-560$\,mA in $20$\,mA steps. (b) The BLS intensity decreases with the increase in current from $-560$\,mA to $-700$\,mA in $20$\,mA steps.
		(c) Time positions of the first (black squares) and the second (red circles) oscillation peak. Grey triangles mark the beginning of the coherent BEC motion after the pumping pulse is switched off. (d) Frequency of the Josephson oscillations (blue triangles) vs applied current in the strip. Vertical bars show 95\% confidence intervals. The measurement error increases for the smallest and highest currents due to the reduced amplitude of the Josephson oscillations. The solid red line is a linear fit by the weighted least squares.
	}
\end{figure}
	
Figure\,\ref{f:2}(e) shows the time evolution of $N\sp{center}(t)$ for the trench geometry with different depths of the trench. We see that the period of the oscillations becomes smaller with the deepening of the trench. To characterize this effect quantitatively, we have determined the explicit time positions of the first and the second peaks, $\tau_1$ and $\tau_2$ [see \Fig{f:5}(a,\,b)], for different depths of the trench and plotted them as black squares and red circle dots in \Fig{f:5}(c). In this figure, grey triangles indicate the zero moment of time, when the pumping pulse is switched off. Starting from this moment the BEC's coherency is established \cite{Kreil2019}, triggering the supercurrent BEC motion \,\cite{Bozhko2019}. Notably, the time interval $\tau_1$ between the zero moment of time and the first Josephson oscillation peak is very close to the time interval between the second and the first oscillation peaks $\tau_2-\tau_1$. The frequency of the oscillations was calculated as the mean value of inverted time intervals $1/\tau_1$, $1/(\tau_2-\tau_1)$ and $2/\tau_2$. It is shown by blue squares in \Fig{f:5}(d). We see that the experimentally found frequency of the oscillations $\Omega\sb{exp}\sp{osc}$ linearly increases with the trench depth $|\Omega_0|$ with the slope $R\sb{exp}\approx 0.023$.

A na\"ive dimensional reasoning results in $\Omega\sb{exp}\sp{osc}\sim\Omega_0$, i.e. $R\sb{exp}$ should be of the order of unity, which is not the case.
We rationalize the observation in the following  Sec.\,III using the solution of the Gross-Pitaevskii equation \cite{Pitaevskii2003, Rezende2010}.
 
It is worth noting that the oscillation character of the BEC evolution  naturally explains the redistribution of the BEC population between the left and right zones over time: for instance, in \Fig{f:3}(b) most of the $+q$-BEC is localized in the right zone, while at a much later time, in \Fig{f:3}(h), the populations in the left and right zones are about the same. This is because during the first half of the Josephson period the right-BEC supercurrent is mostly flowing from the right to the central zone, while during the second half of the period it is flowing from the center zone to the left and to the right zones, although not necessarily by the same amount. As a result, after several Josephson periods, the  occupations of the right- and the left-BECs in the left and right sample zones tend to equilibrium, as indeed observed. As expected, no such an equilibration occurs for the scenario of a potential ridge (\Fig{f:3} Column A). 

It should be noted that in the process of evolution, the density of the condensate decreases and eventually becomes comparable with the thermal noise value. The interaction between the magnon BEC and the gaseous magnons is not considered in our scenario.

\vspace{-3mm}

\section{\label{s:III} III. Josephson oscillations and their discussion}

In this section, we analyze the dynamics of the magnon condensate in the framework of the Gross-Pitaevsky equation, traditionally used to describe BECs \cite{Pitaevskii2003, Rezende2010}. 
\vspace{3mm}
    
\subsection{1. Analytical background}

In order to formulate the Gross-Pitaevskii equations for the magnon BECs, we recall the Holstein-Primakoff transformation in quantum mechanics\,\cite{LL}
\begin{align}
	\begin{split}\label{Sa}
S_+&\=S_x+i S_y=\hbar \sqrt{2s}\ (1- a^\dag a / 2s)^{1/2} a \,  \,,\\
S_-&\=S_x-i S_y=\hbar \sqrt{2s}\ a^\dag(1- a^\dag a / 2s)^{1/2}\,,\\
S_z&\= \hbar (s- a^\dag a)\,, 
    \end{split}
\end{align}
which defines   the spin operators $S_\pm$, $S_z$ in terms of the  boson creation and annihilation operators, $a^\dag$ and $a$, effectively truncating their infinite-dimensional Fock space to the finite-dimensional subspaces. Its classical analogue
\begin{align}
	\begin{split}\label{Ma}
M_+(\bm{r}, t)&\=M_x+i M_y= b \sqrt{\gamma \left( 2M_0 - \gamma b^* b \right)} \,,\\
M_-(\bm{r}, t)&\=M_x-i M_y =M_+^*\,, \\
M_z(\bm{r}, t)&\=M_0 - \gamma b^* b\,,
	\end{split}
\end{align}
represents the magnetization vector $\bm{M}(\bm{r}, t)$ in terms of the canonical variables, the complex spin-wave amplitudes $b(\bm{r}, t)$ and $b^*(\bm{r}, t)$, which are the classical analogues of $a$ and $a^\dag$, see, e.g., Eq.\,(3.4.8) in Ref.\,\cite{Lvov1994}. Here, $M_0$ is the saturation magnetization in the absence of the spin waves and $^*$ denotes complex conjugation. 
	
In order to obtain a simplified  description of two narrow packages of magnons around $\bm{q}=\pm \bm{q}_0$ and $\omega_{\bm{q}}=\omega \sb{min}$, we consider for simplicity only one spatial dimension $x$ and introduce so-called slow-envelope variables
\begin{equation}
B_\pm(x,t)=   \exp( i \omega \sb{min}t)   \!\! \int\! b_{q}(t)\exp \Big[i(q\pm q_0) \cdot x]\frac{d q}{2\pi}\,,
\end{equation}
where $b_{q}(t)$ is the Fourier transform of $b(x,t)$. Then, using the standard procedure (see, e.g., Sec.\,1.5.1. in \cite{ZLF}), one derives from the canonical equations of motion for $b_{q}(t)$ 
\begin{equation}\label{MO}
 i \frac{d b_{q}(t)}{ d t}= \omega_q b_{q}(t) + \mbox{interaction terms} 
\end{equation} 
\noindent
two coupled one-dimensional Gross-Pitaevskii Equations (GPEs) for the $\pm q $-BECs with wave functions $B_\pm(x,t)$ in the external potential $P(x)=\Omega_0\, f(x)$ [given by \Eq{gamma}]:
 	\begin{align}\label{GPE-B}\begin{split}
		i \frac{\partial B_\pm}{\partial t} =& \Big[- \frac{\omega''_{xx}}{2}\frac{\partial^2}{\partial x^2} + \Omega_0\,f(x) \\ & + T|B_\pm|^2   + S |B_\mp|^2 -i\,\Gamma \Big] B_\pm \ .
\end{split}	\end{align} 
Here  $T$ is the amplitude of the self-interaction, $S$ is the cross-interaction amplitude between $+q$- and $-q$-BECs \cite{Dzyapko2017}, $\omega''_{xx}= \partial^2 \omega(\bm{q})/(\partial q_x)^2$ for $\bm{q}=\pm \bm{q}_\mathrm{min}$. In our experiments $\omega''_{xx}\approx 0.7\,$cm$^2$/s.  $\Gamma$ is the damping frequency. 

Introducing new variables that compensate the time decay
 \begin{equation}\label{CB}
    C_\pm(x,t)= B_\pm(x,t) \exp ( \Gamma t)\,,
 \end{equation}
we obtain from \Eq{GPE-B}:
  \begin{align} \label{GPE}
    \begin{split}
        i \frac{\partial C_\pm}{\partial t} =& \Big\{ - \frac{\omega''_{xx}}{2}\frac{\partial^2}{\partial x^2} + \Omega_0\,f(x)\\ & +\Big [  T|C_\pm|^2   + S |C_\mp|^2\Big ] \exp(-2 \Gamma t) \Big\} C_\pm \ .
    \end{split}	
 \end{align}	

In the linear approximation ($T=S=0$), the evolution of the two BEC's $B_\pm(x,t)$ decomposes into two independent processes: the evolution of the stationary condensates $C_\pm( x,t)$ according to \Eqs{GPE} and their exponential decay according to \Eq{CB}. In our experiments, the non-linearity is small and can be peacefully neglected. Indeed, in the whole range of the used magnetic fields, $T\ll |S|$. Moreover, the depth $|\Omega_0|$ of the trench for the currents $|I| \geq 400$\,mA is above $2\pi\cdot 64\,$MHz, while the maximal nonlinear frequency shift (at the moment of switching off the pumping) ${\Omega\Sb{NL}}_\pm =S|B_\pm|^2\simeq 2\pi \cdot 9\,$MHz $\ll |\Omega_0|$ and decays by about two orders of magnitude during the observation time. 

Therefore, we can follow the evolution of the stationary BEC's $C_\pm( x,t)$ by experimentally observing the decaying condensates $B_\pm(x,t)$, provided that the sensitivity of the BLS-system is high enough to observe their evolution during the later stage of the decay. \looseness=-1

\vspace{-6mm}
\subsection{2. Numerical analysis of the magnon Josephson oscillations}

\begin{figure*}  
	\includegraphics[width=1\textwidth]{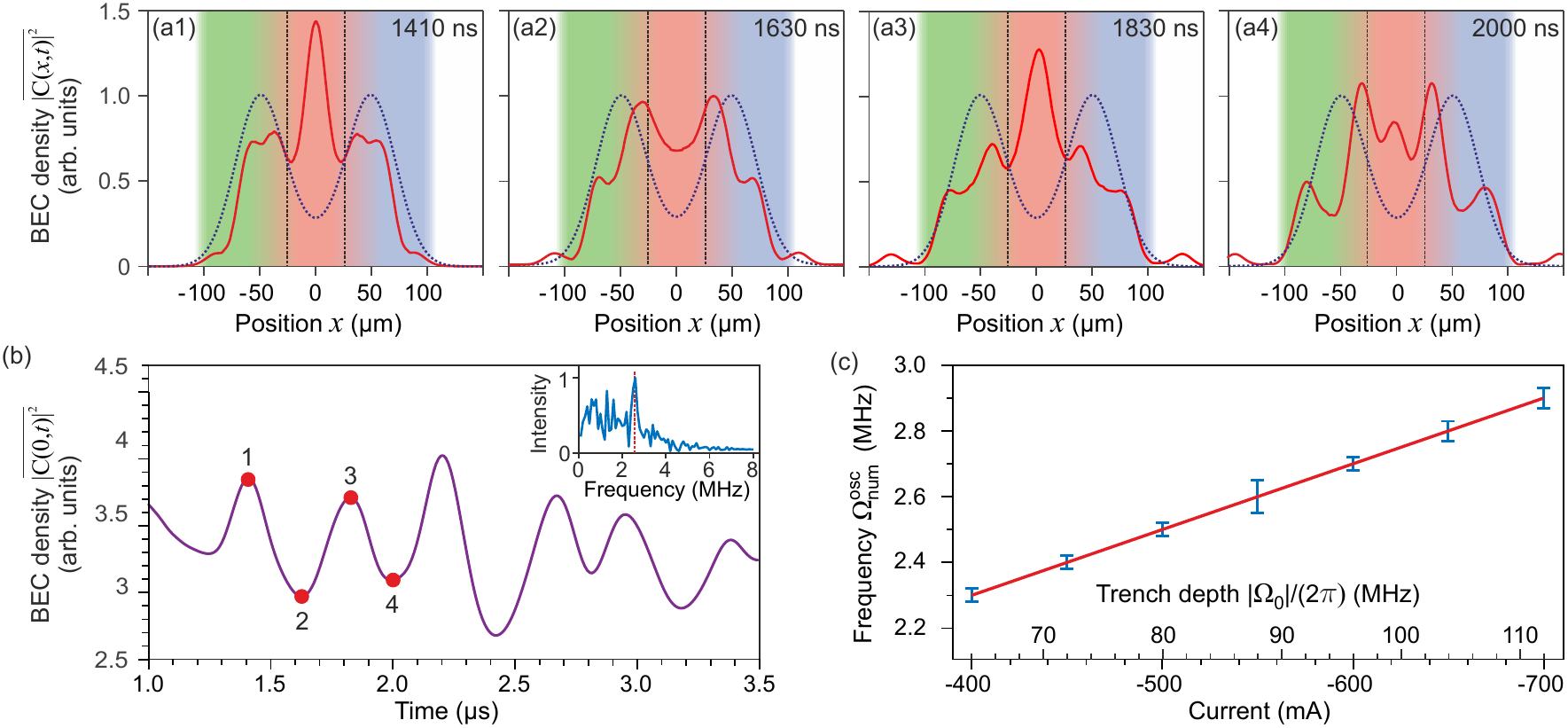}
	\caption{
		\label{f:6} 
		Numerical solutions of \Eq{GPE} with parameters and initial conditions taken from the experiment. The data in panels (a) and (b) were obtained for a trench depth of 80\,MHz, corresponding to a bias current $I= -500\,\mathrm{mA}$. The solid red lines in panels (a1)--(a4) show the spatial distribution of the BEC density $\overline{|{C}(x,t)|^2}$ for four consecutive moments of time, marked by circles in panel (b). The dotted blue lines indicate the initial BEC distribution at zero moment of time. For a better comparison with the experimental data, the BEC distributions were smoothed by a Gaussian filter with a width matching the diameter of the focal spot of the probing laser beam. The vertical dashed lines and shaded colored areas mark the potential trench and different areas of the sample, similar to \Fig{f:1}(b) and \Fig{f:3}.
		Panel (b) plots the BEC density $\overline{|{C}(0,t)|^2}$ in the trench as a function of time. The inset displays the corresponding Fourier spectrum. The dominant spectral peak with a frequency of 2.5\,MHz is marked with a dashed red line.
		Panel (c) shows the dependence of the Josephson frequency $\Omega\sp{osc}\sb{num}$ on the trench depth $|\Omega_0|$. The error-bar values refer to the widths of the corresponding Fourier spectral lines.
		\looseness=-1	
	}
\end{figure*}

To clarify the evolution of the BECs, we numerically solved \Eq{GPE} with parameter values close to the experimental ones, taking symmetric initial conditions shown by the dashed blue lines in \Fig{f:6}(a). To be able to observe a longer evolution of the BEC densities than that observed in the experiment, we put $\Gamma=0$. For simplicity, we assume that $C_+=C_-=C$.

To suppress variations of the BEC density $|C(x,t)|^2$ with a spatial scale of the order of the BEC coherence length $\sim 3\,\mu$m \cite{Bozhko2019}, which cannot be resolved in our experiments, we smoothed the obtained density profiles with a sliding Gaussian filter $\exp[- (x-x_0)^2/(2 \sigma^2)]$. Its dispersion $\sigma=13\,\mu$m matches the diameter of the BLS laser spot. Examples of the smoothed density profiles $\overline{|{C}(x,t)|^2}$ are shown in \Fig{f:6} in the upper panels (a1)--(a4) for $I=500\,$mA corresponding to a trench depth of 80\,MHz.
 
To characterize the BEC evolution in the central zone, we integrated $|C(x,t)|^2$ with the same Gaussian weight $\exp[- x^2/(2 \sigma^2)]$ and yield $ N(t) \=\overline{|{C}(0,t)|^2}$. The evolution of $N(t)$ after some transient time is shown in \Fig{f:6}(b). We clearly see almost periodic variations of $N(t)$. Here, we have marked four characteristic moments in time, two of which ($t_1$ and $t_3$) correspond to times when $N(t)$ approaches local maxima, and the other two ($t_2$ and $t_4$) correspond to times with local minima of $N(t)$. The BEC density profiles shown in Fig.\,6(a1)--(a4) correspond to these instants.

For times $t_1$ and $t_3$, we see pronounced maxima of $\overline{|C (x,t)|^2}$ in the central zone, while the profiles shown in \Figs{f:6}(a2) and (a4) for times $t_1$ and $t_3$ show pronounced dips in this zone. Note that GPE conserves the total number of magnons. This means that the area under the red colored profiles on all panels is the same and is equal to the area under the blue dotted lines indicating the initial conditions. In other words, the entire evolution of the system is an almost periodic redistribution of the magnon BEC density between the central and the side zones, which we interpret as the Josephson oscillations.

To find the frequency of these oscillations, we performed a Fourier analysis of $ N(t)$. An example of the frequency power spectrum is shown in the inset in \Fig{f:6}(b). It has a sharp peak indicated by the vertical red line. The frequency of this peak corresponds to the mean period of the time dependence in \Fig{f:6}(b). The dependence of the frequency of this peak on the depth of the trench is shown in \Fig{f:6}(c). As in the experiment, the numerically found oscillation frequency $\Omega\sp{osc}\sb{num}$ depends linearly on the trench depth $|\Omega_0|$.

The slope of the linear dependence in Fig.\,\ref{f:6}(c) is $R\sb{num}\approx 0.0125$. 
Recall that the experimentally found slope $R\sb{exp}$ is about 0.023. Both slopes are much less than unity, although $R\sb{exp} > R\sb{num}$. The difference in the frequency of oscillations in the experiment and in the numerical analysis can be caused both by the influence of gaseous magnons, not considered in our numerical analysis, and by the fact that we solved the one-dimensional GPE\,\eqref{GPE}, which greatly simplifies the description of the real physical system.

Given the aforementioned simplifications of the analytical description of the problem, we consider the achieved semi-quantitative agreement between experiment and calculations to be quite acceptable.

\vspace{3mm} 
\subsection{3. Analytical scenario of the Josephson oscillations}
  
To rationalize analytically the Josephson oscillations, observed in experiment and numerical simulations, in a manner similar to the description of the  traditional Josephson junction, we neglect the spatial dependence of the magnetic field $\bm{h}$ in the left, right, and central zones and introduce boundary zones of a width $\Delta$, in which the magnetic field varies linearly. These boundary zones play the role of the Josephson junctions. As we notice,  in a na\"{i}ve picture, the frequency of the Josephson oscillations of the supercurrent through the junction should be equal to the magnetic trench depth $|\Omega_0|$. However, the localized BEC packages in the side zones ``hardly know'' about the depth  of the trench in the center of the central zone, where $x=0$.   Being centered at some $x_{\pm}$ away from the central zone, they sense only the local value of the potential $P(x_{\pm})$ and its local slope $d P(x)/ dx$ at $x_{\pm}$. The package can sense the trench depth only by a linear extrapolation $ P(0)\simeq P(x_\pm) - x_\pm\, {d P(x_\pm)}/{d x_\pm}$. 
Consequently, the time evolution of the package 
is expected to be governed by the effective depth of the trench $\Omega_0\sp{eff}$  : 
\begin{equation}\label{osc}
	    	\Omega\sp{eff}_0(x) =x \,\frac{d P(x)}{d x} \= R\sp{eff} (x) \Omega_0\ ,
		\end{equation} 
		instead of $\Omega_0$.

To estimate in this scenario the ratio $R \sp{eff}(x)= \Omega_0 (x)/ \Omega_0$ we find the BEC peak positions in \Fig{f:3} as $x \simeq 70\,\mu$m. Then, using the potential profile in \Fig{f:2}(b), we calculate $d P(x)/dx$ and estimate $R\sp{eff}(x)$ to be about 0.089. This value is essentially smaller than unity and not far away from the experimentally found ratio $R\sb{exp}\approx 0.023$. We consider this agreement as a support of the suggested simple analytical scenario of the Josephson oscillations.

\vspace{6mm}
\section{IV.  Summary }

The discovery of AC Josephson oscillations in the system of two room-temperature magnon condensates opens the path for a better understanding of the underlying physics of this phenomenon. For instance, the observation of these oscillations clearly demonstrates the ability of the magnon supercurrent to accumulate phase shifts multiple times larger than $2\pi$, what was sometimes questioned before \cite{Sonin2017,Sonin2020}. Furthermore, we identify several directions where further research is needed. The first one is the study of the evolution of the BEC spatial profiles $N(x,t)$ for the case of a time-dependent magnetic trench/ridge, which can be achieved by applying time-dependent dc~currents. Secondly, it will be especially interesting to  realize a conventional ``superconductor geometry'' (see, e.g., Ref.\,\onlinecite{Nakata2014}), considering two weakly coupled magnon systems with a controlled BEC phase shift between them.

The numerical solutions of the GPE\,\eqref{GPE}, supported by the simple analytical scenario, describes the main observed phenomena and explains the discovered Josephson oscillations as a consequence of the potential energy difference between the BECs in the induced magnetic trench and in the nearby regions. Nevertheless, the achieved level of understanding of the open physical phenomenon still requires further elaboration and detailing. For example, one needs to account for the interactions of the bottom gaseous magnons with the BEC magnons by formulating a model analogous to the two-fluid model of superfluid helium \cite{Landau}. 

We believe that these findings will pave the way to various engineering applications in spatially-inhomogeneous setups, such as information processing in perspective magnon spintronic devices \cite{Nakata2014, Chumak2015, Byrnes2012, Andrianov2014}. 

\section{Acknowledgements}
Financial support of the European Research Council within the Advanced Grant 694709 SuperMagnonics -- ``Supercurrents of Magnon Condensates for Advanced Magnonics'' as well as financial support of the Deutsche Forschungsgemeinschaft (DFG, German Research Foundation) through the Collaborative Research Center ``Spin+X: Spin in its collective environment'' TRR -- 173 -- 268565370 (project B04) is gratefully acknowledged. We thank V.\,S.\,Tyberkevych for fruitful discussions.\\

\vspace{-6mm}
\section{Appendix: Experimental Methods}

\vspace{-2mm}	
\subsection{1. Sample}
\vspace{-3mm}
	The Yttrium Iron Garnet (YIG, $\mathrm{Y_{3}Fe_{5}O_{12}}$) \cite{Cherepanov1993} sample is 3.5\,mm long and 5\,mm wide. The single-crystal YIG film \cite{Glass1988} of 5.1\,$\mu$m thickness has been grown in the (111) crystallographic plane on a $500\,\mathrm{\mu m}$-thick Gadolinium Gallium Garnet (GGG, $\mathrm{Gd_{3}Ga_{5}O_{12}}$) substrate by liquid-phase epitaxy at the Department of Crystal Physics and Technology of the Scientific Research Company ``Electron-Carat'' (Lviv, Ukraine).
	To ensure a uniform reflectivity of the probing light from the sample under investigation, regardless of the underlying reflectivity of any kind of microstrip structure or its substrate material, a dielectric mirror has been sputtered onto the YIG-film surface. The mirror itself consist of several silicon and titanium oxide layers with a total thickness of approximately $1\,\mu$m.

\vspace{-4mm}	
\subsection{2. Experimental setup: details}
\vspace{-3mm}
	A sketch of the experimental setup is shown in Fig.\,\ref{f:2}(a). It consists of microwave, DC and optical parts. 
	
	The microwave part is used as the source for the pumping pulses. It includes a microwave generator and a power amplifier connected to a microstrip pumping circuit through a bias tee. In the presented experiments, the pulse duration is 300\,ns, the pulse repetition rate is 3\,kHz, and the carrier frequency is 13\,GHz. The $50\,\mathrm{\mu m}$-wide grounded microstrip line, fabricated on top of a dielectric substrate (RT/duroid$^\circledR$ 6010), is used to induce the pumping microwave magnetic field. 
	The multilayer sample comprising a dielectric mirror, a YIG film, and a GGG substrate is placed on top of the microstrip so that the dielectric mirror faces the microstrip near its grounded end in the region of the maximum microwave magnetic field.
	Here, a microstrip pump resonator, conventionally used in magnon BEC experiments \cite{Demokritov2006, Bozhko2016}, is replaced by a grounded microstrip to transmit both microwave and pulsed DC currents. Since the quality factor of such a microstrip is unity, the applied microwave pumping power $P_\mathrm{pump}$, required to reach the threshold of the Bose--Einstein condensation is tens of times higher than in resonator experiments. It is about $400\,\mathrm{W}$.
    To ensure that no undesirable heating effects affect the measurements, we used sufficiently short pump pulses separated by a long time interval used to relax the sample. It allows for enough time for the pumped magnons to decay and for the generated heat to dissipate into the surrounding space. 
	
	The DC part, which is decoupled from the microwave circuit by a bias tee, feeds the microstrip with a pulsed DC current (the pulse duration is 5\,$\mu$s and the pulse repetition rate is 3\,kHz) in the range of $\pm 700$\,mA to create the positive or negative magnetic profiles of different depths in the pumping area. This current is pulsed to reduce any disturbing additional heat load. The relative frequency shift of the bottom frequency $\omega_\mathrm{min}$ calculated for the longitudinal component $h_x$ of the DC magnetic field created by the microstrip at a distance $d=1\,\mu$m (equal to the thickness of the sputtered mirror) is shown in Fig.\,\ref{f:2}(b). To calculate this magnetic field, we assumed a uniform current distribution in a rectangular microstrip cross section of $50\,\mu$m width and $17\,\mu$m height \cite{phdchumakov.antennafield}. 
	
	The optical part [Fig.\,\ref{f:2}(a)] is used for the probing of the magnon Bose--Einstein condensate (BEC) by means of Brillouin light scattering (BLS) spectroscopy. Its main parts are the probing green laser (single-mode, $532\,\mathrm{nm}$ wavelength) and a multipath tandem Fabry--P\'erot interferometer. 
	The probing beam of power $0.5\,\mathrm{mW}$ is focused onto the sample surface into a focal spot of about $50\,\mathrm{\mu m}$ in diameter. The scattered light is directed to a multipass tandem Fabry--P\'{e}rot interferometer \cite{Sandercock1981,Mock1987,Hillebrands1999} for frequency selection with resolution of 100\,MHz. A single photon counting avalanche diode detector is placed at the output of the interferometer. The output of the detector is then connected to a counter module synchronized with a sequence of microwave pulses. Every time the detector registers a photon, this event is recorded with a time stamp to a database that collects the number of photons ensuring a time resolution of 250\,ps. The frequency of the interferometer transmission is also recorded, thus providing frequency information for each detected photon. \looseness=-1
	
	The small power of the probing light of 0.5\,mW and the low electrical resistance of the current-conducting microstrip of 0.3\,Ohm in combination with the chosen pulsed regimes for microwave and DC currents ensure a rather small heat deposition to the YIG sample (the calculated temperature increase does not exceed 0.15\,K) and, thus, a negligible contribution of heating effects to the observed phenomena. This fact is confirmed by the BEC dynamics shown in Fig.\,\ref{f:2}(d), where no heat influence (e.g., in the form of enhanced BEC decay reported in Refs.\,\cite{Bozhko2016,Bozhko2019,Kreil2018,Kreil2019}) is observed throughout the entire range of applied DC currents.
	
	Spatially-resolved probing of the magnon dynamics is performed by the controlled displacement of the sample using a precise linear positioning stage. The entire stage is placed directly between the poles of the electromagnet, ensuring high field uniformity and stability. The setup allows for space-resolved measurements of the magnon dynamics in the YIG-film across the current-conducting microstrip as it is shown in Fig.\,\ref{f:2}. The scanning step value is set to $20\,\mathrm{\mu m}$.

\vspace{-4mm}	
\subsection{3. Frequency- and wavevector-selective Brillouin light scattering spectroscopy}
\vspace{-3mm}
	Brillouin light scattering can be understood as a change in the frequency $\omega_{_\mathrm{L}}$ and wavevector $\bm{q}_{_\mathrm{L}}$ of photons, when they annihilate or create quanta of magnetic collective excitations, namely magnons. In our case, a probing laser light beam illuminates the YIG-film sample at a certain incident angle $\Theta$  laying in the $(\bm{\^x}, \bm{\^z})$ plane. In order to detect the inelastically scattered photons, they need to travel the same path back [see Fig.\,\ref{f:2}(a)]. This can happen according to the following scheme. The out-of-plane wavevector of the incident photon is inverted due to its reflection from a dielectric mirror deposited on the semi-transparent YIG film. The inversion of the in-plane photon's wavevector is possible if the photon creates an in-plane magnon with the frequency $\omega_\mathrm{m}$ and wavenumber $q_x = 2q_\mathrm{_L} \sin(\Theta)$ (Stokes process: the frequency of the scattered photon is $\omega'_\mathrm{_L} = \omega_\mathrm{_L}-\omega_\mathrm{m}$), or if an in-plane magnon with the opposite wavenumber $q_x = -2q_\mathrm{_L} \sin(\Theta)$ transfers its momentum to the incident photon and is annihilated (anti-Stokes process: $\omega'_\mathrm{_L} = \omega_\mathrm{_L} + \omega_\mathrm{m}$). The frequency of the inelastically scattered photons is then analyzed by a Fabry--P\'{e}rot interferometer and the intensities of the Stokes and anti-Stokes spectral peaks are related to the densities of magnons with different sign in their wavenumbers $q_x$, namely $+q_\mathrm{min}$ and $-q_\mathrm{min}$ in our case. 
	Note that due to the contribution to the scattering of both linear and quadratic coefficients of magneto-optical coupling, there is a difference in the Stokes and anti-Stokes scattering intensities depending on the polarization direction of the light electric field relative to the $\textit{\textbf{H}}$ magnetic field \cite{Sandercock1973}. In our experiment, this leads to a consistently lower measured BLS intensity af the $-q$-BEC compared to the $+q$-BEC (see Fig.\,\ref{f:3}).
	\looseness=-1

\vspace{-4mm}	
\subsection{4. Processing of experimental data} 
\vspace{-3mm}
	To obtain the dependencies of the Josephson oscillation's frequency on the magnitude of the DC current, the following experimental data processing was performed. The first step was to reduce the noise level in the raw experimental data, which are exemplarily presented in Figs.\,\ref{f:2}(c,\,e). A bilateral frequency filter, based on the combination of two Gauss filters, was applied to reach this goal. This type of filtering provides edge-preserving noise-reducing smoothing \cite{bilateral}. Secondly, the decay compensation was performed to avoid the artificial shifting of maxima (Josephson peaks) on experimentally acquired BLS waveforms ($\propto|B_\pm|^2$ in Eq.\,4) towards shorter times. For this compensation, we used the decay constant $\Gamma=2.5$\,MHz defined from the decrease of the BLS signal measured for $I_\mathrm{dc}=0$. The obtained waveforms, which are proportional to $|C_\pm|^2$ in Eq.\,7a, are shown in Figs.\,\ref{f:5}(a,\,b).
	
	The next step was to determine the time positions $\tau_1$ and $\tau_2$ of the first and the second Josephson peaks [see Figs.\,\ref{f:5}(a,\,b)] relative to the moment when the pumping was turned off, and the BEC began to move. First, each of the peaks was cropped from the bottom at half of its height to reduce the distortion of the peak shapes due to the overlapping with other peaks. Afterwards, the peak time positions were calculated as the first moments of the functions representing the upper peak parts divided by the respective zeroes moments of these functions. This approach significantly increases the immunity of the obtained results to noise-induced random variations in the peak shapes.
	
	The frequency of the Josephson oscillations for the given DC current was calculated as the mean value of inverted time intervals $1/\tau_1$, $1/(\tau_2-\tau_1)$ and $2/\tau_2$. Considering the increase of the measurement error for the lowest and largest DC currents, the linear fit of the frequency dependence [red line in Fig.\,\ref{f:5}(d)] was calculated using the weighted least squares method.
	
%

\end{document}